\documentclass[english]{article}
\usepackage[margin=1.0in]{geometry}
\usepackage[T1]{fontenc}
\usepackage[latin9]{inputenc}
\usepackage{mathrsfs}
\usepackage{amstext}
\usepackage{amsmath}
\usepackage{amssymb}
\usepackage{graphicx}
\usepackage{caption}
\usepackage{subcaption}
\usepackage{authblk}
\usepackage{color, soul}
\def \apj {{\it Astrophys.~J.}}
\def \aj {{\it Astronom.~J.}}

\def \prd {{\it Phys.~Rev.~D}}
\def \prl {{\it Phys.~Rev.~Lett.}}
\def \mnras {{\it Mon. Not. R.~Astr.~Soc.}}
\def \nature {{\it Nature}}

\def \AnA {{\it Astr. Astrophys.}}
\def \grg {{\it Gen. Rel. Grav.}}
\def \pl {{\it Phys. Lett.}}

\makeatletter
\providecommand{\tabularnewline}{\\}

\makeatother

\usepackage{babel}
\begin{document}

\title{\textbf{Variable Chaplygin Gas: Constraints from Supernovae, BAO, Look Back Time, and GRBs}}
\author[a]{Bhuvan Agrawal\thanks{abhuvan053@gmail.com}}
\author[b]{ Geetanjali Sethi\thanks{getsethi@gmail.com}}
\author[b]{Shruti Thakur}
\affil[a]{\it\small Department of Applied Physics, Delhi Technological University}
\affil[b]{\it\small Department of Physics, St. Stephen's College, University of Delhi}
\maketitle

\begin{abstract}
We examine the variable Chaplygin gas (VCG) model with the aim of establishing tight constraints by using type-Ia supernovae, lookback time measurements, Baryon Acoustic Oscillations (BAO) and the Gamma Ray Bursts. We report the parameter constraints obtained via statistical analysis of cosmic observables namely, luminosity distance, BAO distance and lookback time measurement.
\end{abstract}

\section{Introduction}

Bolstering our understanding of the Universe, observations also have the potential to revolutionize the way we perceive our reality. Two decades back one such observation discovered that the Universe is expanding. From the observations of distant supernovae type-Ia (SNe-Ia) by two independent groups \cite{a, b} it has been established  that the expansion is accelerating. This fact is corroborated by a number of separate observations such as Baryon Acoustic Oscillations (BAO), Cosmic Microwave Background, growth of structures and Gamma-ray Bursts (GRBs).

The dynamics of the Universe is governed by the Einstein's general theory of relativity, but it is not possible to explain the accelerated expansion if only the observable (normal) matter is taken into consideration. Different cosmological models have been proposed to explain our observations. Among the various explanations, the one that is prevalent and most favourable is the $\Lambda$-Cold Dark Matter ($\Lambda$CDM) model, where $\Lambda$ represents the cosmological constant. In this model, $\Lambda$ accounts for the vacuum energy or the energy density of space. However this model suffers from serious fine tuning problems \cite{e, f}. Hence many more models have been explored in the past and this remains an active area of research. Different models explored can be broadly categorized into two classes. One class of models involve reforming the geometry part of the Einstein's equations. This includes generalisation of gravity action and higher dimension spacetime. The other class of models alter the matter component of the Universe in the Einstein's equations. In this approach exotic matter with negative pressure is added to the mass distribution of the Universe. Some of the models based on this are quintessence, k-essence, tachyons, barotropic fluid etc. These are collectively known as Dark Energy models. Recently, an alternative class of models have been proposed which involve a slowly evolving and spatially homogenous scalar field \cite{ratra, zlatev} or two coupled fields \cite{santos}. However, these "Quintessence'' models also suffer from fine tuning problem.

Quintessence, k-essence are scalar field models, while Dark Energy models include barotropic fluids whose pressure is a function of energy density, $P=f(\rho)$. The relationship between pressure and energy density determines the dynamics of the fluid. One such example of a barotropic fluid is the Chaplygin gas \cite{kamen, billic}. By introducing a cosmic fluid in lieu of the cosmological constant and quintessence one can unify the CDM and the $\Lambda$ models' features into a single component with an exotic equation of state. This versatility of the models built on the Chaplygin gas becomes an attractive feature as these models can explain both dark energy and dark matter with a single component. The equation of state for the Chaplygin gas is $P=-A/\rho$, where A is a positive constant. A more {\it generalized} model of Chaplygin gas is characterised by an equation of state
\begin{equation}
P_{ch}=-\frac{A}{\rho_{ch}^{\alpha}}
\end{equation}

where $\alpha$ is a constant such that $0<\alpha\leq1$. Generalised Chaplygin gas automatically leads to an asymptotic accelerated expanding Universe. We see that the normal Chaplygin gas model is assumed for $\alpha =1$. By using the energy-momentum conservation, $d(\rho a^{3}) = -pd(a^{3})$, we can see that the equation of state evolves as \cite{bento}
\begin{equation}
\rho{}_{ch} = \bigg( A+\frac{B}{a^{3(1+\alpha)}} \bigg) ^ {\frac{1}{1+\alpha}}
\end{equation}

where $a$ is the scale factor, and $B$ is the constant of integration. From the above equation it can be seen that at early times, $a\ll1$,  we have $\rho \propto a^{-3}$ and it behaves as CDM. For late times, $a\gg1$, and we get $p = -\rho = constant$, i.e. the case of the cosmological constant which leads to the observed accelerated expansion. The models based on Chaplygin gas have garnered great interest as they have shown promise in the past. These models have been found to be consistent with the SNe Ia data \cite{makler}, CMB peak locations \cite{bento1} and other observational tests like gravitational lensing, cosmic age of old high redshift objects etc. \cite{dev}, as also with some combination of some of them \cite{bean}. Chaplygin gas models can also be accommodated within the standard structure formation scenarios \cite{billic,bento,fabris}. Therefore, the Chaplygin gas model seems to be a good alternative to explain the accelerated expansion of the universe. However the Chaplygin gas model produces oscillations or an exponential blowup of matter power spectrums that are inconsistent with observations \cite{teg}.

A modification of Chaplygin gas recently proposed \cite{cgas} is {\it variable} Chaplygin gas. The model was constrained using SNe Ia 'gold' data \cite{g}. In this paper, we obtain better constrains for the variable Chaplygin gas model parameters using statistical analysis of SNe Ia in Union 2.1 compilation from the Supernova Cosmology Project \cite{scp}, Baryon Acoustic Oscillation (BAO) from the Sloan Digital Sky Survey (SDSS), lookback time measurements, and the GRBs. The basic formalism of the model is reviewed in Section 2. Section 3 gives the general theory for cosmic observables employed for the statistical analysis - luminosity distance, look back time and baryonic acoustic oscillations. Analysis of datasets and results are presented in Section 4. We discuss the results and conclusions in Section 5.

\section{Variable Chaplygin Gas Model}
\subsection{Theory}
In this section, we consider the {\it variable} Chaplygin Gas (VCG) \cite{cgas}, which is characterised by the equation of state:
\begin{equation}
P{}_{ch}=-\frac{A(a)}{\rho_{ch}}
\label{eq:equationofstate}
\end{equation}

where $A(a)=A_{0}a^{-n}$ is a positive function of the cosmological scale factor $a$. $A_{0}$ and $n$ are constants. Using the energy conservation equation, $d(\rho a^{3}) = -pd(a^{3})$,  in a flat Friedmann-Robertson-Walker universe and equation (\ref{eq:equationofstate}), the variable Chaplygin gas density evolves as:
\begin{equation}
\rho{}_{ch}=\sqrt{\frac{6}{6-n}\frac{A{}_{0}}{a{}^{n}}+\frac{B}{a{}^{6}}}
\end{equation}

where B is a constant of integration. For $n=0$, the original Chaplygin gas behaviour is restored. The gas initially behaves as dust-like matter ($\rho_{ch} \propto a^{-3}$) and later as a cosmological constant ($p = -\rho = constant$). However, in the present case the Chaplygin gas evolves from dust dominated epoch to cosmological constant in present times (see \cite{cgas}).

The Friedmann equation gives the expansion rate of the Universe in terms of matter and radiation density, $\rho$, curvature, $k$, and the cosmological constant, $\Lambda$, as
\begin{equation}
H^{2} \equiv \bigg( \frac{\dot{a}}{a}\bigg)^{2} = \frac{8\pi G}{3}\rho-\frac{k}{a^{2}}+\frac{\Lambda}{3}
\end{equation}

After reducing it to the case of the spatially flat Universe,
\begin{equation}
H{}^{2}=\frac{8\pi G}{3}\rho
\end{equation}

where $H\equiv \dot{a}/a$ is the Hubble parameter. Therefore the acceleration condition $\ddot{a} > 0$ is equivalent to
\begin{equation}
\left(3-\frac{6}{6-n}\right)a^{6-n}>\frac{B}{A_{0}}
\end{equation}

In order to incorporate accelerated expansion of the Universe, the necessary condition is $n<4$. This gives the present value of energy density of the variable Chaplygin gas 
\begin{equation}
\rho_{cho}=\sqrt{\frac{6}{6-n}A_{0}+B}
\end{equation}

where $a_{0}=1$. Defining a parameter, $\Omega_{m}$,
\begin{equation}
\Omega_{m}=\frac{B}{6A_{0}/(6-n)+B}
\end{equation}

the energy density becomes 
\begin{equation}
\rho_{ch}(a)=\rho_{ch0}\left[\frac{\Omega_{m}}{a^{6}}+\frac{1-\Omega_{m}}{a^{n}}\right]^{1/2}
\label{eq:energydensity}
\end{equation}

We also use the deceleration parameter, $q=-(\ddot{a}a/\dot{a}^{2})$. It is a measure of the cosmic acceleration and expansion of space, for an accelerated expansion $q<0$. Now, the general expression for deceleration parameter in terms of redshift, $z$, and Hubble parameter, $H$, is given by
\begin{equation}
q(z)=\frac{d\ln H(z)}{d\ln (1+z)}-1
\end{equation}

In addition to deceleration parameter, calculation of the age of the Universe is also employed in order to draw a comparison between the evolution of the Universe with redshift in the cases of LCDM and VCG models. The age of the universe is given by
\begin{equation}
t(z)=\int_{z}^{\infty}\frac{dz'}{(1+z')H(z')}
\end{equation}

\subsection{Model}
The Friedmann equation, using equation (\ref{eq:energydensity}) for a variable Chalygin gas, becomes
\begin{equation}
H^{2}=\frac{8\pi G}{3}\bigg\{\rho_{r0}(1+z)^{4}+\rho_{b0}(1+z)^{3}+\rho_{ch0}\Big[\Omega_{m}(1+z)^{6}+(1-\Omega_{m})(1+z)^{n}\Big]^{1/2}\bigg\}
\label{eq:friedmanvcg}
\end{equation}

where $\rho_{r0}$ and $\rho_{b0}$ are the present values of energy densities of radiation and baryons, respectively. Using\footnote{We have used the fact that for a flat Universe, $\Omega_{b0}+\Omega_{r0}+\Omega_{ch0}=1$, i.e the total matter density sums up to unity.}
\begin{equation}
\frac{\rho_{r0}}{\rho_{ch0}}=\frac{\Omega_{r0}}{\Omega_{ch0}}=\frac{\Omega_{r0}}{1-\Omega_{r0}-\Omega_{b0}}
\end{equation}

and
\begin{equation}
\frac{\rho_{b0}}{\rho_{ch0}}=\frac{\Omega_{b0}}{\Omega_{ch0}}=\frac{\Omega_{b0}}{1-\Omega_{r0}-\Omega_{b0}},
\end{equation}

Equation (\ref{eq:friedmanvcg}) becomes,
\begin{equation}
H^{2}=\Omega_{ch0}H_{0}^{2}a^{-4}X^{2}(a),
\end{equation}

where
\begin{equation}
X^{2}(a)=\frac{\Omega_{r0}}{1-\Omega_{r0}-\Omega_{b0}}+\frac{\Omega_{b0}a}{1-\Omega_{r0}-\Omega_{b0}}+a^{4}\bigg(\frac{\Omega_{m}}{a^{6}}+\frac{1-\Omega_{m}}{a^{n}}\bigg)^{1/2}.
\end{equation}

\section{Cosmological Observables}
In this work, we analyse three different cosmological  observables - Luminosity Distance, Lookback Time, and BAO distance measurement. As we will see in this section, these observables show an explicit dependence on the cosmological model under consideration. These parameters are chosen in order to compare the experimentally obtained and the theoretical values of the respective model parameters.

\subsection{Luminosity Distance}
We have used the SNe-Ia to constrain the parameters of the variable Chaplygin gas model. Using the Friedmann equation, in a flat Universe, the luminosity distance is expressed as
\begin{equation}
d_{L}(z, \mathbf{p})=c(1+z)\int_{0}^{z} \frac{dz'}{H(z',\mathbf{p})}
\label{eq:ld_rel}
\end{equation}

where, $\{\mathbf{p}\}$ denotes the set of all parameters describing the cosmological model and $H(z, \mathbf{p})$ denotes the Hubble parameter as defined in the model chosen. While calculating, we have taken into consideration the contributions from radiation and baryons in addition to the Chaplygin gas. The distance modulus is obtained from the above relation using
\begin{equation}
\mu_{th}=5\log\frac{H_{0}d_{L}}{ch}+42.38
\end{equation}

$H_{0}$ being the value of Hubble parameter at present, $c$ is the speed of light and $\mathit{h}\equiv H_{0} / 100 km s^{-1} Mpc^{-1}$ is the dimensionless Hubble parameter. The theoretical distance modulus from the equation above is compared to the observed values from the dataset. To determine the best fit parameters we minimise
\begin{equation}
\chi^{2}=\sum_{i}\bigg[\frac{\mu_{th}^{i}-\mu_{obs}^{i}}{\sigma_{i}}\bigg]^{2}-\frac{C_{1}}{C_{2}}\bigg(C_{1}+\frac{2}{5}\ln10\bigg)-2\ln\mathit{h},
\end{equation}

here, 
\begin{equation}
C_{1}\equiv\sum_{i}\frac{\mu_{th}^{i}-\mu_{obs}^{i}}{\sigma_{i}^{2}},
\end{equation}
\begin{equation}
C_{2}\equiv\sum_{i}\frac{1}{\sigma_{i}^{2}}.
\end{equation}

In conjunction with the Union 2.1 compilation \cite{scp}, we also incorporate GRBs data sample as described in ref. \cite{grb}. Gamma-ray bursts are short-lived violent explosions that release high intensity gamma radiation. These are the most energetic events in the cosmos, they have the capacity to deliver more energy in a few seconds than that our Sun will emit in its entire 10 billion year lifetime. For this reason, they are detectable upto high redshifts and are used to study the nature of the Universe and thus the dark energy.

\begin{figure}
\centering
\includegraphics[scale=0.75]{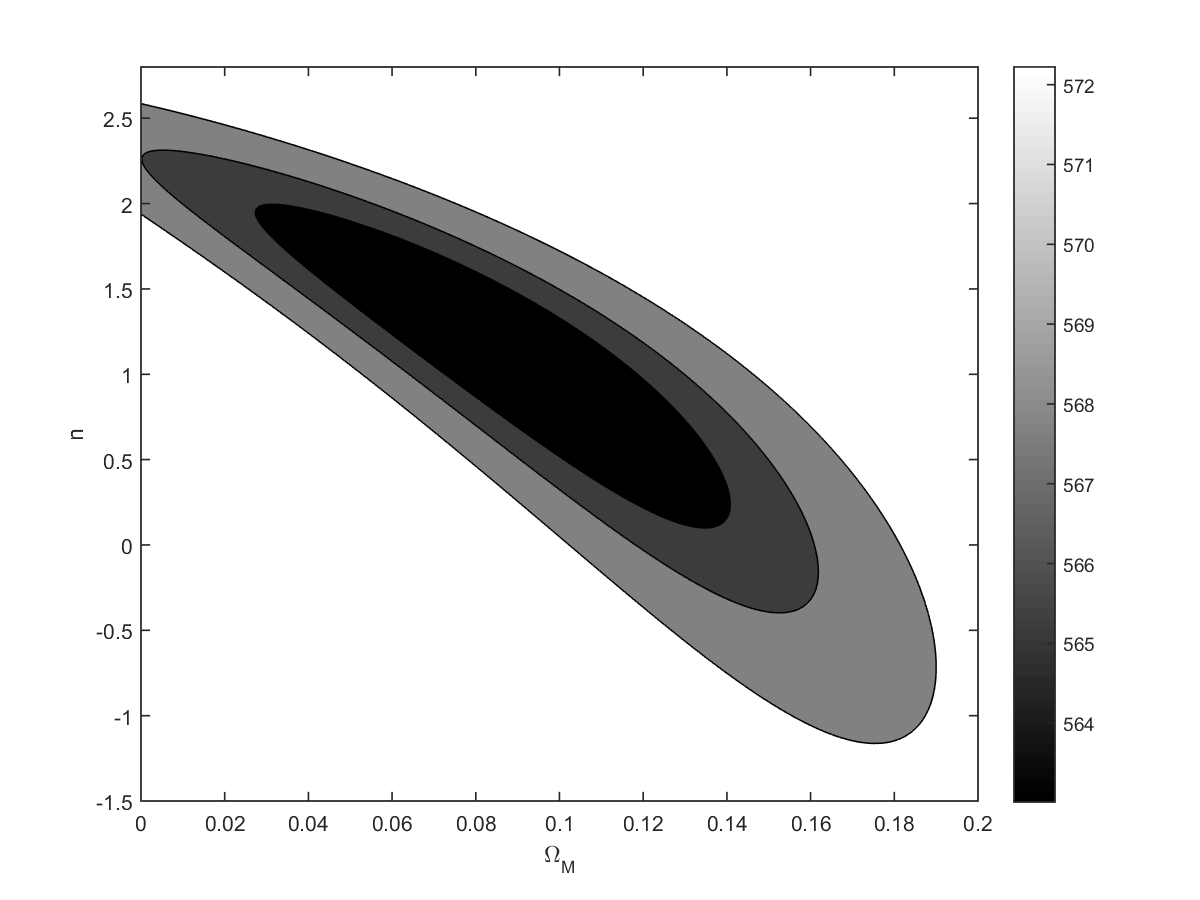}
\caption{\it $\chi^{2}$ contours for Union 2.1 compilation on $\Omega_{m}-n$ parameter space.}
\label{fig:LD1}
\end{figure}

\begin{figure}
\centering
\includegraphics[scale=0.75]{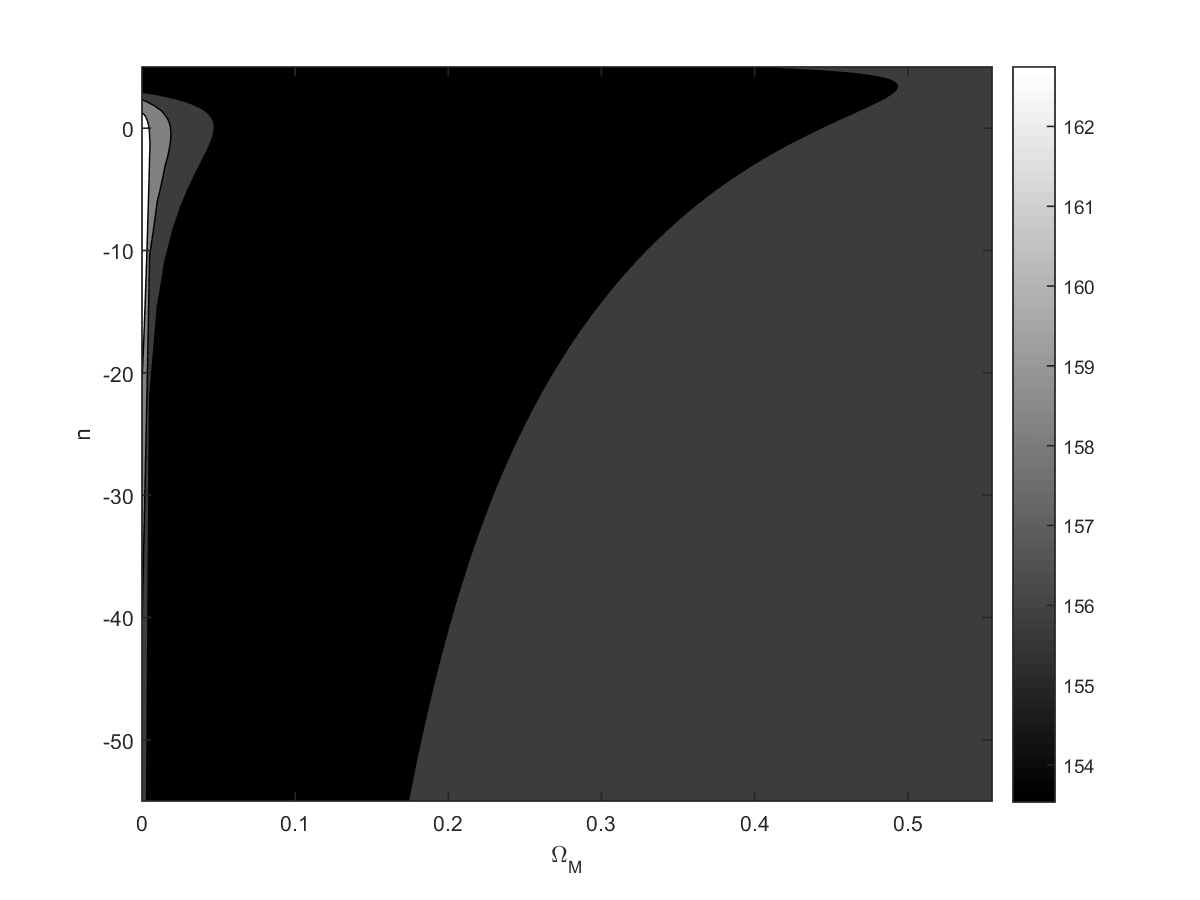}
\caption{\it $\chi^{2}$ contours for GRB dataset on $\Omega_{m}-n$ parameter space.}
\label{fig:grb}
\end{figure}

\subsection{Look Back Time}
We consider a homogeneous, isotropic, spatially flat universe described by the Friedmann-Robertson-Walker metric with the line element $ds{}^{2}=dt^{2}-a^{2}(t)(dx^{2}+dy^{2}+dz^{2})$, where $a(t)$ is the cosmological scale factor.

The lookback time is defined as the difference between the age of the Universe today, $t_{0}$, and its age, $t_{z}$, at redshift $z$. The dependence of lookback time on redshift z can be written as
\begin{equation} \label{eq:lbt}
t_{L}(z,\mathbf{p})=H_{0}^{-1}\int _{0}^{z} \frac{dz'}{(1+z')\mathscr{H}(z', \mathbf{p})}
\end{equation}

where $\mathscr{H}(z, \mathbf{p})=H(z)/H_{0}$ denotes the dimensionless Hubble parameter as defined in the model chosen. In order to calculate the age of the Universe, we use equation (\ref{eq:lbt}) with one small change - the upper limit of integration is increased from $z$ to infinity.	

To use the lookback time measurements, consider an object $i$ at redshift $z$. The age of the object is defined as the difference between the age of the Universe when the object was born, i.e. at the formation redshift $z_{F}$, and the one at $z$. That is,
\begin{equation}
\begin{split}
t_{i}(z)&=\int_{z}^{\infty}\frac{dz'}{(1+z')\mathscr{H}(z', \mathbf{p})}-\int_{z_{F}}^{\infty}\frac{dz'}{(1+z')\mathscr{H}(z', \mathbf{p})} \\
&=\int_{z}^{z_{F}}\frac{dz'}{(1+z')\mathscr{H}(z', \mathbf{p})} \\
&=t_{L}(z_{F})-t_{L}(z)
\end{split}
\end{equation}

here, we have used the equation (\ref{eq:lbt}) for lookback time. This relation is extended to all the $N$ objects. Using this relation, we calculate the lookback time $t_{L}^{obs}(z_{i})$ as
\begin{equation}
\begin{split}
t_{L}^{obs}(z_{i})&=t_{L}(z_{F})-t_{i}(z) \\
&=[t_{0}^{obs}-t_{i}(z)]-[t_{0}^{obs}-t_{L}(z_{F})] \\
&=t_{0}^{obs}-t_{i}(z)-df
\end{split}
\end{equation}

where $t_{0}^{obs}$ is the estimated age of the Universe today, and a {\it delay factor} can be defined as 
\begin{equation}
df=t_{0}^{obs}-t_{L}(z_{F})
\end{equation}

We introduce an incubation time or delay factor in order to account for the formation redshift $z_{F}$, i.e. the amount of time since the beginning of the structure formation in the Universe until the formation time of the object. In this work, we have used the dataset as described in \cite{samushia} in addition with \cite{dantas}. In this case as well, we minimize
\begin{equation}
\chi^{2}=\sum_{i=1}^{N}\bigg[\frac{t_{L}^{th}(z_{i}, \mathbf{p})-t_{L}^{obs}(z_{i})}{\sqrt{\sigma_{i}^{2}+\sigma_{t}^{2}}}\bigg]^{2}+\bigg(\frac{t_{0}^{th}(\mathbf{p})-t_{0}^{obs}}{\sigma_{t_{0}^{obs}}}\bigg)^{2}+\bigg(\frac{\mathit{h}-\mathit{h}^{obs}}{\sigma_{\mathit{h}}}\bigg)^{2},
\label{eq:lbtchi}
\end{equation}

Now, to analytically obtain marginalised chi square, we define a log-likelihood function as
\begin{equation}
\tilde{\chi}^{2}=-2ln\int_{0}^{\infty}d\tau\exp\bigg(-\frac{1}{2}\chi_{age}^{2}\bigg)
\end{equation}
 
where $\tau$ represents the delay factor, $\sigma_{t}$ is the uncertainty on observed age of the Universe, $t_{0}^{obs}$, and $\sigma_{i}$ is that on $t_{L}^{obs}(z_{i})$. The modified log-likelihood function can be rewritten as, 
\begin{equation}
\tilde{\chi}^{2}=A-\frac{B^{2}}{C}+D-2ln\bigg[\sqrt{\frac{\pi}{2C}}erfc\bigg(\frac{B}{2C}\bigg)\bigg]
\end{equation}

here,
\begin{equation}
A=\sum_{i=1}^{N}\frac{\Delta^{2}}{\sigma_{i}^{2}+\sigma_{t}^{2}}, \;\;\;\;\;\;\;\;\;\;\;\; B=\sum_{i=1}^{N}\frac{\Delta}{\sigma_{i}^{2}+\sigma_{t}^{2}}, \;\;\;\;\;\;\;\;\;\;\;\; C=\sum_{i=1}^{N}\frac{1}{\sigma_{i}^{2}+\sigma_{t}^{2}},
\end{equation}

$\Delta=t_{L}^{th}-[t_{0}^{obs}-t_{L}^{obs}]$, and $D$ is the sum of last two terms in equation (\ref{eq:lbtchi}).
We use an estimate of $t_{0}^{obs}$, for which we choose $(t_{0}^{obs}, \sigma_{t})=(13.6, 0.35) Gyr$ \cite{dantas}. Also, $\mathit{h}^{obs}$ is the estimate value of $\mathit{h}$ with $\sigma_{\mathit{h}}$ its uncertainty, $(h, \sigma_{h})=(0.72, 0.08)$ following from the Hubble space telescope key project results \cite{t}.

\begin{figure}
\centering
\includegraphics[scale=0.75]{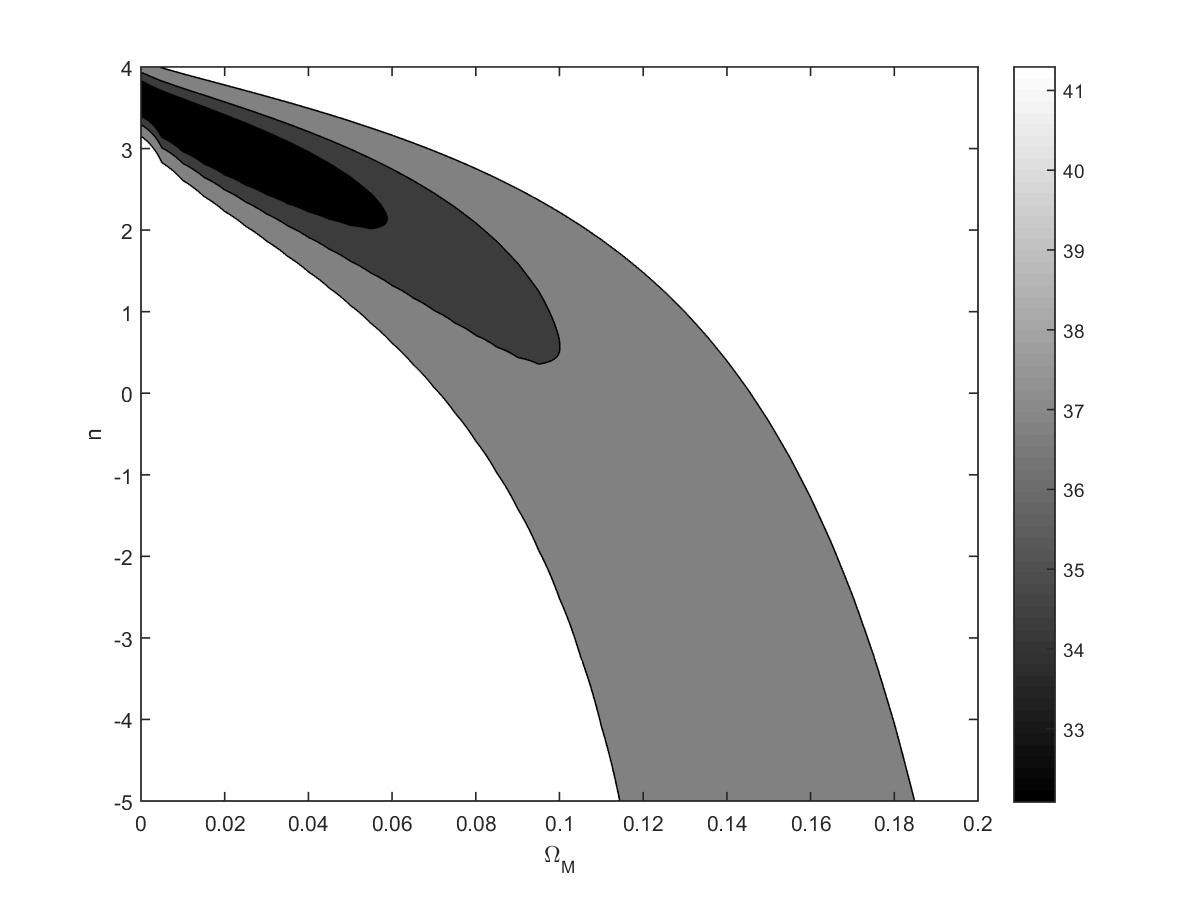}
\caption{\it $\chi^{2}$ contours from lookback time measurement.}
\label{fig:LBT1}
\end{figure}

\subsection{Baryonic Acoustic Oscillations}
Photons in the early Universe were trapped in the hot, dense plasma of electrons and baryons (protons and neutrons). Being scattered by plasma via Compton scattering, photons were restricted to a considerably low mean-free path, up until the Universe cooled down to low enough temperatures that the electrons and protons could combine to form Hydrogen ({\it Recombination}). Prior to recombination, photons and matter are remain coupled together. In an overdense region, the photon and matter oscillate together under the influence of both gravitational attraction and outward radiation pressure. Once the photons decouple from the neutral matter, its density distribution gets imprinted on the radiation spectrum. These, now free, photons are observed in the Cosmic Microwave Background (CMB).

These density fluctuations are the result of acoustic density waves in the ionised matter in the early Universe. The perturbations of baryons starts to grow and interact with the dark matter perturbations leaving an impression in the large-scale structure formation at late time \cite{i}. BAO matter clustering can thus be used as{\it 'Standard Ruler'} \cite{j}. Table \ref{Tab:bao_data} contains the BAO distance measurements used.

\begin{table}
\centering
\begin{tabular}{|c|ccc|}
\hline
$Survey$ & $z$ & $d_{z}(z)$ & $Reference$\tabularnewline
\hline
\hline
$6dFGS$ & $0.106$ & $0.3360 \pm 0.0150$ & $\cite{k}$\tabularnewline
$MGS$ & $0.15$ & $0.2239 \pm 0.0084$ & $\cite{l}$\tabularnewline
$BOSS LOWZ$ & $0.32$ & $0.1181 \pm 0.0024$ & $\cite{m}$\tabularnewline
$SDSS(R)$ & $0.35$ & $0.1126 \pm 0.0022$ & $\cite{n}$\tabularnewline
$BOSS CMAS$ & $0.57$ & $0.0726 \pm 0.0007$ & $\cite{m}$\tabularnewline
$WiggleZ$ & $0.44$ & $0.073$ & $\cite{o}$\tabularnewline
$WiggleZ$ & $0.6$ & $0.0726$ & $\cite{o}$\tabularnewline
$WiggleZ$ & $0.73$ & $0.0592$ & $\cite{o}$\tabularnewline
\hline 
\end{tabular} 
\protect\caption{BAO distance measurements}
\label{Tab:bao_data}
\end{table}

The distance-redshift relation for BAO measurement is given by
\begin{equation}
d_{z}=\frac{r_{s}(z_{drag})}{D_{V}(z)},
\end{equation}

where
\begin{equation}
r_{s}(z_{drag})=\frac{c}{\sqrt{3}}\int_{z_{drag}}^{\infty}\frac{dz}{H(z)\sqrt{1+(3\Omega_{b0}/4\Omega_{r0}(1+z)^{-1}}}
\label{eq:comsnhr}
\end{equation}

is the radius of the comoving sound horizon at the drag epoch $z_{drag}$ i.e. when photons and baryons decouple \cite{p}. Also,
\begin{equation}
D_{V}(z)=\bigg[\frac{czd_{C}^{2}(z)}{H(z)}\bigg]^{1/3}
\end{equation}

is the volume-averaged distance \cite{q} and $d_{C}(z)=d_{L}(z)/(1+z)$ is the comoving angular diameter distance, with $d_{L}(z)$ being the luminosity distance given by equation (\ref{eq:ld_rel}). $\Omega_{b0}$ and $\Omega_{r0}$ in equation (\ref{eq:comsnhr}) are the present values of baryon and photon density parameters receptively. Throughout our calculations, the dimensionless Hubble parameter is chosen to be $h=0.71$, also we have taken $\Omega_{b0}=0.05$ and $\Omega_{\gamma0}=9.89\times10^{-5}$ as in \cite{r}.

The three measurements from the WiggleZ survey in table \ref{Tab:bao_data} are correlated. The covariance matrix included while using these data points is as follows.
\[
C^{-1}=
\begin{pmatrix}
1040.3 & -807.5 & 336.8 \\
-807.5 & 3720.3 & -1551.9 \\
336.8 & -1551.9 & 2914.9
\end{pmatrix}
\]

For each of the survey in table \ref{Tab:bao_data}, except the WiggleZ survey, chi-square is given as
\begin{equation}
\chi^{2}=\bigg[\frac{d_{z}^{th}-d_{z}^{obs}(z, \mathbf{p})}{\sigma}\bigg]^{2}
\end{equation}

For the WiggleZ survey, we use the correlation matrix and the chi-square is written as
\begin{equation}
\chi_{WiggleZ}^{2}=[d_{z, i}^{th}-d_{z}^{obs}(\mathbf{p})]^{T}C^{-1}[d_{z, i}^{th}-d_{z}^{obs}(\mathbf{p})].
\end{equation}

The calculated $\chi^{2}$ values are then added together in order to obtain the joint chi-square value for the complete BAO distance measurement.

\begin{figure}
\centering
\includegraphics[scale=0.75]{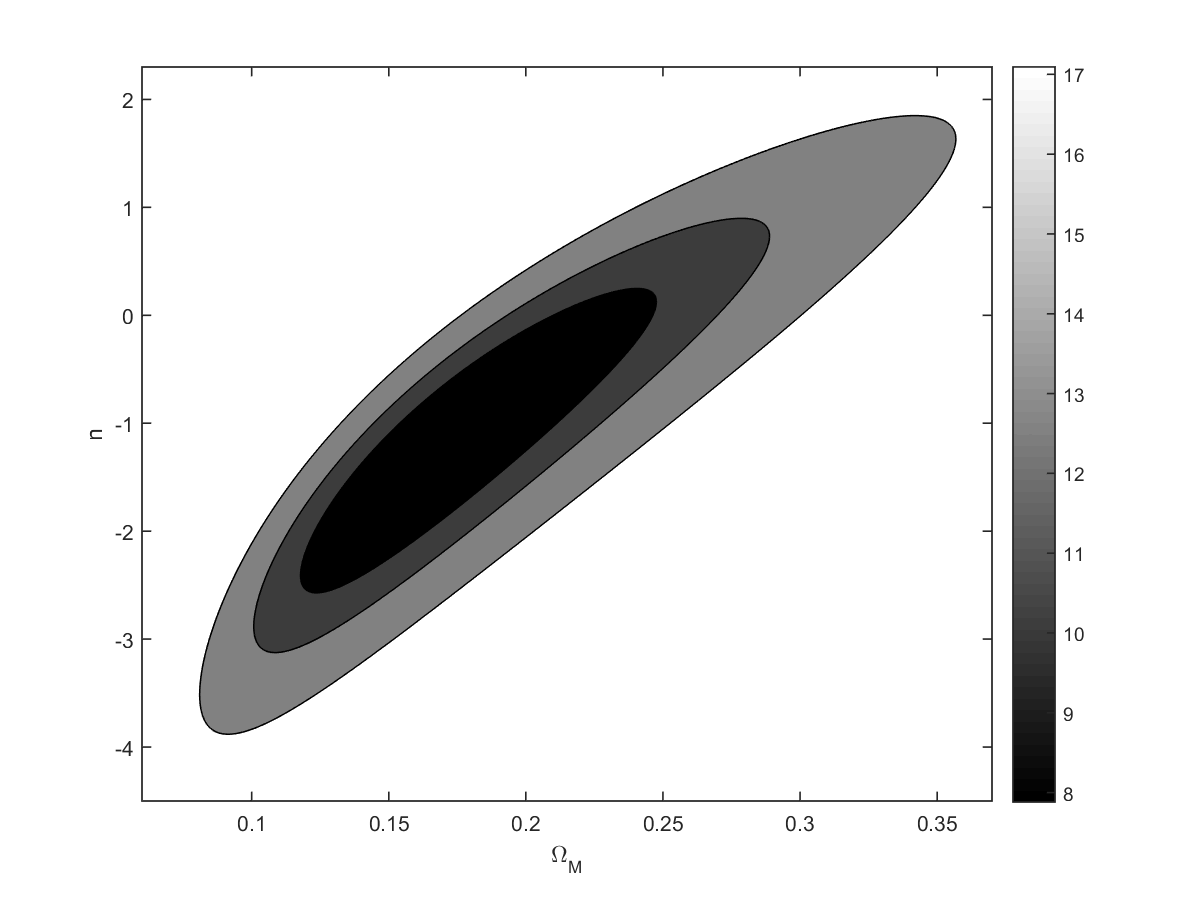}
\caption{\it $\chi^{2}$ contours from BAO distance measurement}
\label{fig:BAO1}
\end{figure}

\section{Statistical Analysis and Results}
\begin{figure}
\centering
\includegraphics[scale=0.75]{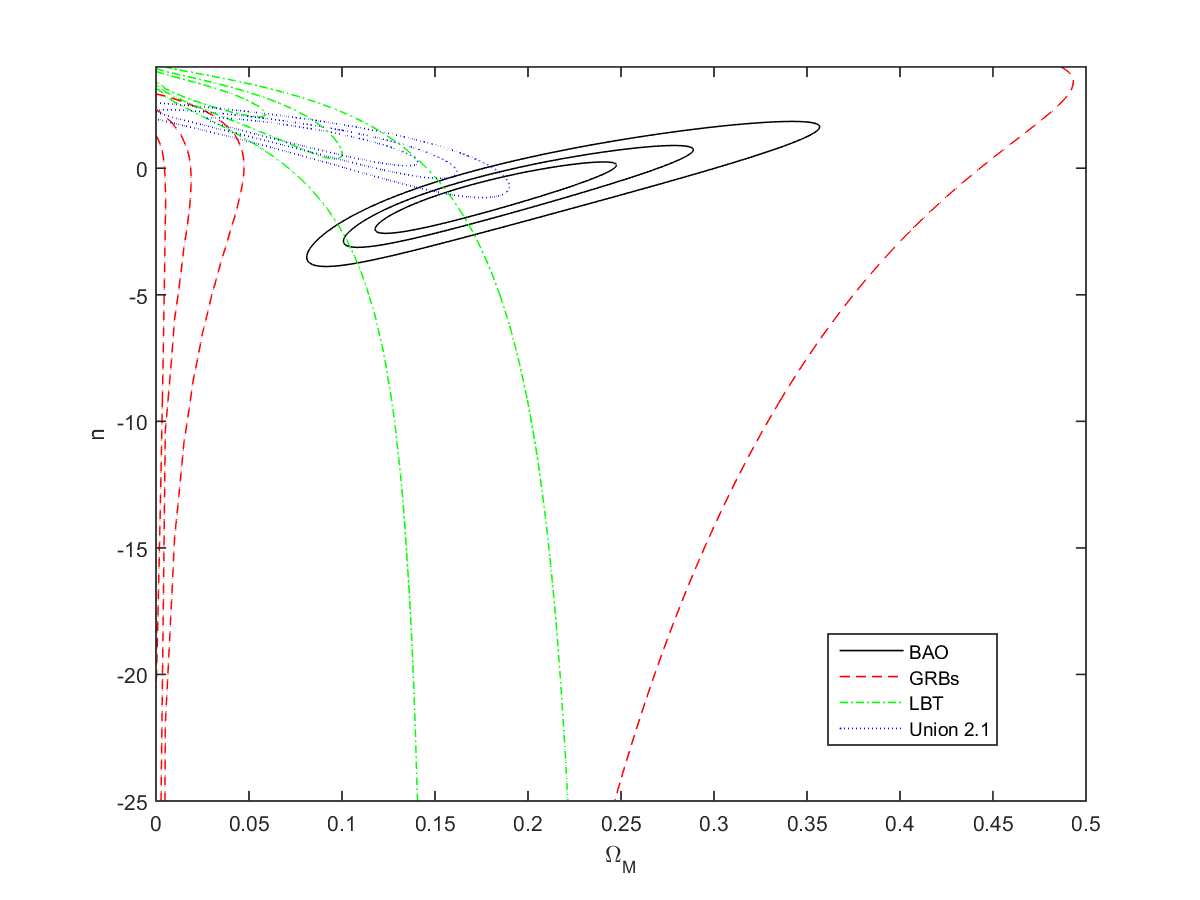}
\caption{\it $\chi^{2}$ contours from Union + Lookback time + BAO distance + GRBs}
\label{fig:combined}
\end{figure}

In order to constrain the parameters of the variable Chaplygin model we subject the model to statistical analysis using the four data sets and parameters described in the previous section. From figure \ref{fig:combined}, it is clear that contours of the three observables result in an overlap of the $1\sigma$ confidence levels.

To estimate the best fit to the parameters the likelihood function is defined as
\begin{equation}
\mathscr{L}\propto \mathrm{exp}[-\chi^{2}(z;\mathbf{p})/2],
\end{equation}

where the $\chi^{2}$ functions for all the datasets are calculated using equations described in their respective subsections. The parameter set, $\mathbf{p}$, contains all the parameter including the{\it nuisance} parameters. These parameters are the dimensionless Hubble parameter, $\mathit{h}$, the observed age of the Universe, $t_{0}^{obs}$, and the delay factor, $df$, and the observed age of the Universe, $t_{0}^{obs}$ are marginalised over. We maximise the marginalised likelihood given by

\begin{equation}
\mathscr{L}_{p_{i}}\propto\int dp_{1} ... \int dp_{i-1} \int dp_{i+1} ... \int dp_{n} \mathscr{L}(\mathbf{p})
\end{equation}

After marginalisation these likelihood values are normalised and we obtain the contour plots. From the Union 2.1 compilation, $\chi_{min}^{2} = 563.0114$ with $579$ degrees of freedom, at $\Omega_{m}=0.087$ and $n=1.14$. As shown in figure \ref{fig:LD1}, the $1\sigma$ constraints are $\Omega_{m}\in[0.0265, 0.1415]$ and $n\in[0.085, 2.01]$. In addition to these, lookback time measurements result in $\chi_{min}^{2}=32.0827$ with $45$ degrees of freedom, and the best fit parameters are $\Omega_{m}=0.01$ and $n=3.29$. Figure \ref{fig:LBT1} gives the range of $\Omega_{m}\in[0, 0.0592]$ and $n\in[-2.01, 3.84]$. Whereas, measurements from the BAO give $\chi_{min}^{2}=7.8819$ with $7$ degrees of freedom, at $\Omega_{m}=0.171$ and $n=-1.21$. Figure \ref{fig:BAO1} gives the parameter constraints as $\Omega_{m}\in[0.117, 0.2485]$ and $n\in[-2.586, 0.265]$. The GRBs were used to obtain the contour plot as shown in figure \ref{fig:grb}. We report that $\chi_{min}^{2}=153.5349$ with $161$ degrees of freedom, at $\Omega_{m}=0.0$ and $n=4.27$. While figure \ref{fig:combined} overlays the three plots. $1\sigma (68.3\%)$, $2\sigma (95.4\%)$, and $3\sigma (99.7\%)$ confidence regions are shown. We see that the best fit parameters obtained in this work are in agreement with those reported in \cite{cgas} which are $\Omega_{m}=0.25$ and $n=-3.4$ with $\chi_{min}^{2}=174.54$.

\begin{figure}
\centering
\includegraphics[scale=0.65]{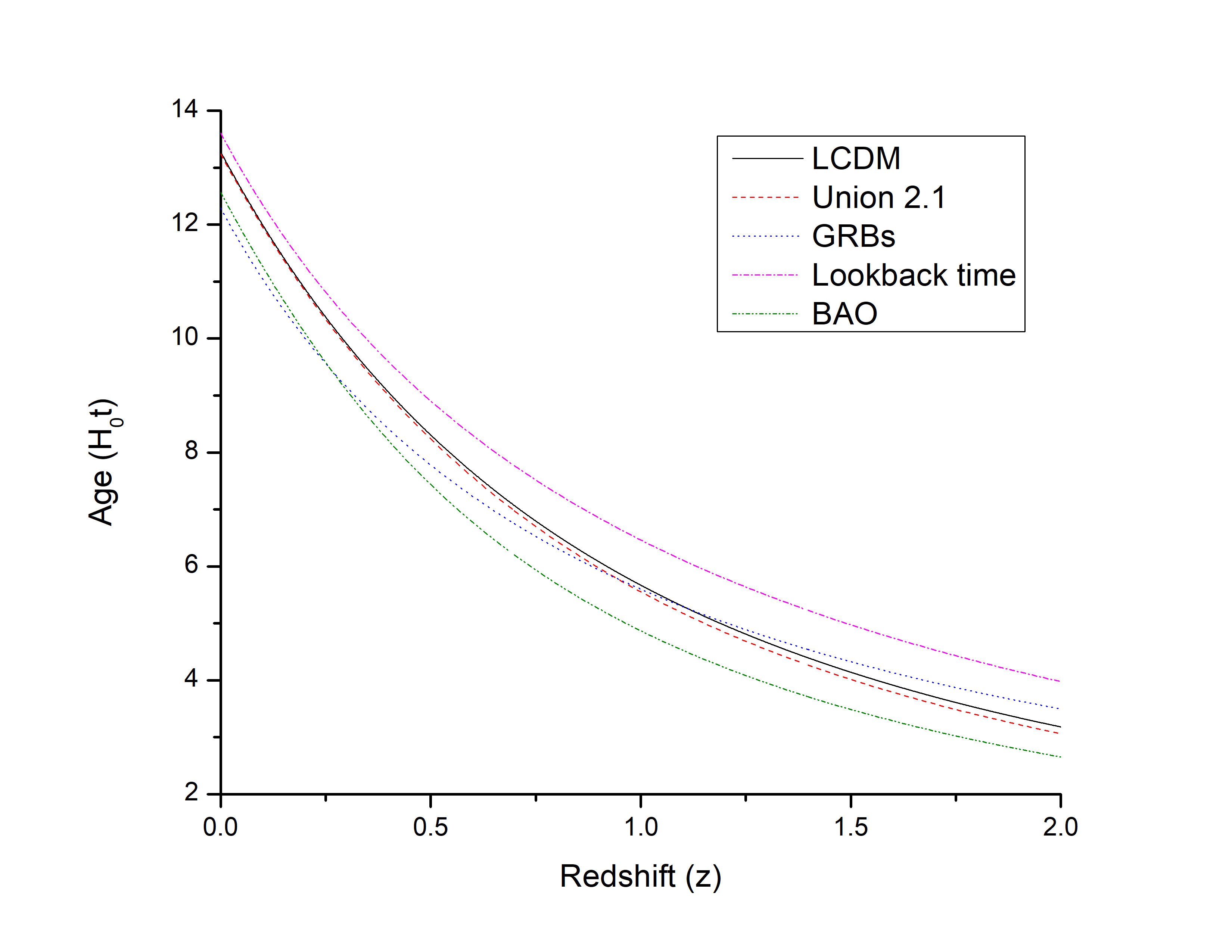}
\caption{\it Age of the Universe, $H_{0}t(z)$}
\label{fig:H0tz}
\end{figure}

\begin{figure}
\centering
\includegraphics[scale=0.65]{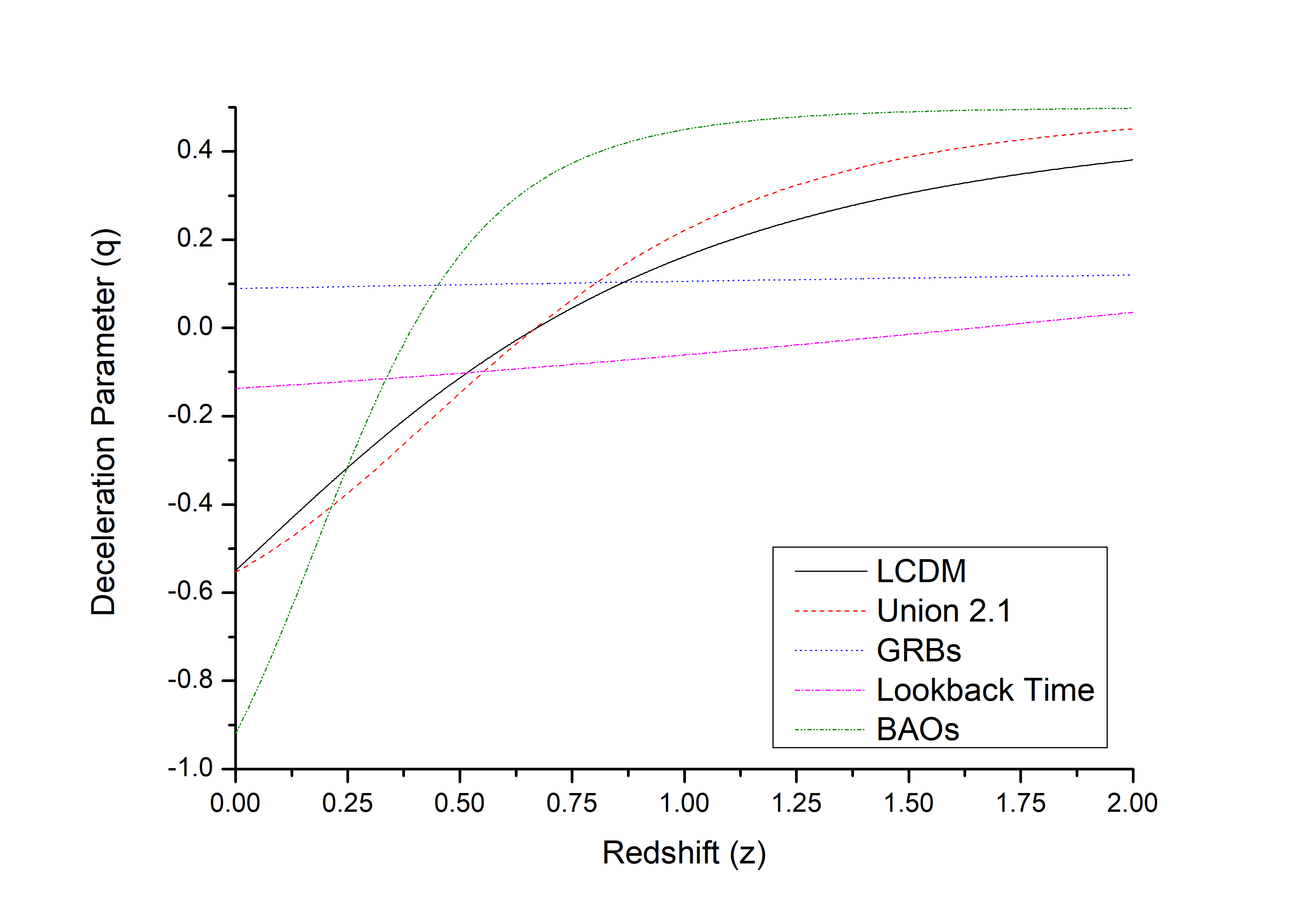}
\caption{\it Deceleration Parameter, $q(z)$}
\label{fig:qz}
\end{figure}

Using the best fit parameters obtained we plot the evolution of age of the universe and deceleration parameter, the best fit values are listed in table \ref{Tab:ageandq}. For the $\Lambda$CDM model, we use $\Omega_{b0}=0.3$ and $\Omega_{\Lambda0}=0.7$. Figure \ref{fig:H0tz} shows the evolution of the age of the universe with redshift, $z$. Figure \ref{fig:qz} shows the evolution of deceleration parameter with redshift. Table \ref{Tab:bounds} shows the bounds obtained for the variable Chaplygin gas model obtained from the four datasets.

\begin{table}
\centering
\begin{tabular}{ccc}
\hline
$Method$ & $t_{0} (Gyr)$ & $q$\tabularnewline
\hline
$Union 2.1$ & $13.243$ & $-0.554$ \tabularnewline
$GRBs$ & $12.291$ & $0.089$ \tabularnewline
$Lookback time$ & $13.608$ & $-0.137$ \tabularnewline
$BAO$ & $12.568$ & $-0.919$ \tabularnewline
$LCDM model$ & $13.277$ & $-0.55$ \tabularnewline
\hline
\end{tabular} 
\protect\caption{Best fit parameter values of age of the Universe and deceleration parameter}
\label{Tab:ageandq}
\end{table}

\begin{table}
\centering
\begin{tabular}{ccc}
\hline
$Dataset$ & $\Omega_{m}$ & $n$\tabularnewline
\hline
$Union 2.1$ & $0.087_{-0.039, -0.082}^{+0.0365, +0.071}$ & $1.14_{-0.676, -1.434}^{+0.586, +1.117}$ \tabularnewline[1.2ex]
$GRBs$ & $0_{-0}^{+0.191}$ & $4.27_{-1.77}^{+0.73}$ \tabularnewline[1.2ex]
$Lookback Time$ & $0.01_{-0.01, -0.01}^{+0.0268, +0.0786}$ & $3.29_{-0.68, -2.356}^{+0.455, +0.636}$ \tabularnewline[1.2ex]
$BAO$ & $0.171_{-0.0382, -0.0675}^{+0.0479, +0.109}$ & $-1.21_{-0.92, -1.81}^{+0.96, +1.98}$
\tabularnewline[1.2ex]
\hline
\end{tabular} 
\protect\caption{$1$ and $2-\sigma$ likelihood bounds on $\Omega_{m}$ and $n$.}
\label{Tab:bounds}
\end{table}

\section{Discussion}
We see that the variable Chaplygin gas model is able to account for the observations regarding the evolution of the universe. Initially the gas behaves like non-relativistic matter and later is able to account for the accelerated expansion of the universe. With our analysis of the Union 2.1 compilation, BAO distance, look-back time measurements and the GRBs, we see that the ranges obtained for $\Omega{}_{m}$ and $n$ not only lie within the confidence levels obtained by Zong-Kuan Guo et al. but we have further constrained the possible values of $\Omega{}_{m}$ and $n$ by a joint analysis of the four datasets and have obtained tight limits on the parameters.

\section{Acknowledgement}
The authors would like to thank Dr. Sampurnanand for their comments and suggestions and The Centre for Theoretical Physics, St. Stephen's College, University of Delhi for facility and support. Dr. Geetanjali Sethi and Dr. Shruti Thakur are grateful to the Principal, St. Stephen's College, University of Delhi for his support.

\end{document}